\documentstyle[multicol,aps,prl,epsfig]{revtex}

\renewcommand{\narrowtext}{\begin{multicols}{2} \global\columnwidth20.5pc}
\renewcommand{\widetext}{\end{multicols} \global\columnwidth42.5pc}

\input{epsf.tex}

\def\top#1{\vskip #1\begin{picture}(290,80)(80,500)\thinlines \put(
65,500){\line( 1, 0){255}}\put(320,500){\line( 0, 1){
5}}\end{picture}}

\begin{document} 
 
\draft 
 
\title{Quantum Breakdown of the Quantized Hall Insulator} 
 
\author{U. Z\"ulicke$^1$ and Efrat Shimshoni$^2$} 
 
\address{$^1$Institut f\"ur Theoretische Festk\"orperphysik, 
Universit\"at Karlsruhe, D-76128 Karlsruhe, Germany} 
 
\address{$^2$Department of Mathematics-Physics, 
Oranim--Haifa University, Tivon 36006, Israel} 
 
\date{\today} 
 
\maketitle 
 
\begin{abstract} 
We present an analytical scaling theory for localization in a  
two--dimensional hierarchical network model that is designed to 
represent {\em phase-coherent\/} electron transport in the quantum-Hall
regime. Scaling expressions for both the longitudinal and Hall
resistivities are derived. In agreement with recent numerical 
studies, we find that the Hall resistivity is quantized in the 
metallic phase but {\em diverges\/} in the insulating phase. This 
suggests that the characteristics of a {\em quantized Hall insulator\/} can
occur only in the presence of a strong dephasing mechanism. 
\end{abstract} 
 
\pacs{PACS number(s): 73.43.Cd, 72.20.My} 
 
\narrowtext 
 
The low--temperature transport properties of disordered 
two--dimensional (2D) electron systems subject to a strong 
perpendicular magnetic field exhibit a multitude of transitions 
between distinct phases\cite{qhe-phas-tran}. These include 
primarily the various quantum Hall (QH) phases, characterized by 
quantized values of the Hall resistivity; $\rho_{\text{H}}=h/e^2 
\nu$ in a wide range of carrier densities and magnetic fields 
centered around certain integer or fractional values of the 
filling factor $\nu$. These plateaus in $\rho_{\text{H}}$ are 
accompanied by a vanishing longitudinal resistivity 
$\rho_{\text{L}}$. At sufficiently high magnetic field, the 
series of QH--to--QH transitions is terminated by a transition to
an insulator, marked by a {\em divergence\/} of $\rho_{\text{L}}$. 
Both kinds of transitions can be associated with competing trends
toward localization and delocalization of electrons in the last,
partially filled Landau level\cite{efrat:prl:97}. 
  
\begin{figure}[b]
\centerline{\epsfig{file=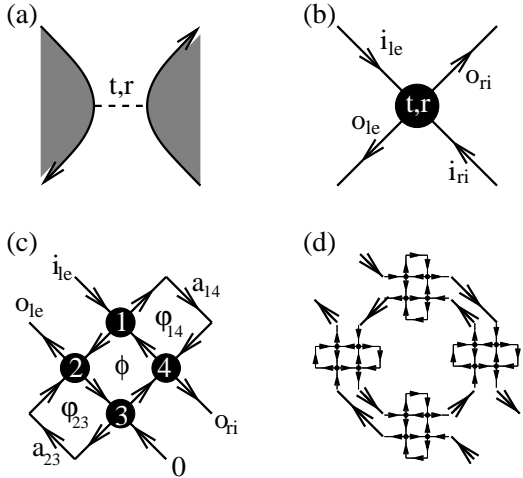,width=2.7in}} 
\vspace{0.3cm} 
\caption{Network model for describing the QH-to-insulator 
transition. (a)~A saddle point between adjacent electron puddles.
Motion within a puddle is directed along chiral edge channels. 
(b)~Saddle point represented by a scattering matrix (vertex) 
relating outgoing electron amplitudes ($o_{\text{le}}, 
o_{\text{ri}}$) to incoming ones ($i_{\text{le}},i_{\text{ri}}$).
(c)~Elementary cell of the hierarchical network. Quantum phases 
acquired by electrons when moving around its three small closed 
loops are indicated. (d)~Lattice at the second level of the 
hierarchy.} 
\label{elemcell} 
\end{figure}
While insulating and conducting phases are distinguished quite 
unambiguously by a diverging or vanishing $\rho_{\text{L}}$, the 
behavior of $\rho_{\text{H}}$ in the insulator can be rather 
subtle. Theoretical studies assuming a low magnetic field $B$ 
indicate a possible dependence on the disorder-averaging 
procedure\cite{HI-lowB}. In particular, certain models find a 
{\em Hall insulator\/} which is characterized by a finite 
$\rho_{\text{H}}\sim B$ similar to the classical (Drude) form. In
the proposed\cite{kiv:prb:92} phase diagram of the QH effects, 
such a Hall insulator is expected to exist in the limits of 
strong disorder or high magnetic field, in agreement with some
experimental results\cite{hi-exp}. More recent
measurements\cite{qhi-exp} have identified an insulating regime
where $\rho_{\text{H}}$ is quantized at the plateau value of the
nearby QH state --- hence dubbed a {\em quantized Hall insulator\/} (QHI).
A quantized $\rho_{\text{H}}$ is consistent with transport models
that assume the existence of a local conductivity 
tensor\cite{ruz:prl:95}. It was later proved\cite{efrat:prb:97} 
on more general grounds that quantization of $\rho_{\text{H}}$ is
a robust feature of an arbitrary network of weakly coupled QH 
puddles at fixed $\nu$, provided transport on the network is  
{\em incoherent\/} (i.e., governed by {\em classical\/} Kirchhoff laws). 
Recent numerical studies\cite{assa:prl:99} indicate that quantum 
interference actually destroys the QHI: once the 
localization length becomes much shorter than the dephasing 
length, $\rho_{\text{H}}$ diverges. Experimental data consistent 
with such a strongly localized regime are not yet available. 
 
In the present work, we examine this intriguing possibility of a 
quantum breakdown of the QHI within an analytic approach, 
designed to gain more insight into the underlying localization 
mechanism. We consider electron transport on a random network 
that is constructed as a hierarchical lattice (see 
Fig.~\ref{elemcell}) and obtain scaling expressions for {\em 
both\/} longitudinal {\em and\/} Hall resistivities. Our 
calculation provides a 2D generalization of the familiar 
scattering approach to 1D 
localization\cite{pwa:prb:80,mello:prb:87,boris:pmb:87}. In 
contrast with the 1D case, we find an unstable fixed point for 
which $\rho_{\text{L}}$ does not scale. It separates a metallic 
phase with quantized $\rho_{\text{H}}$ from an insulating one 
where $\rho_{\text{H}}$ diverges exponentially with system size.

Our approach is inspired by the extensive use of network 
models\cite{chalk:jpc:88} as a framework for studying transport 
in QH samples. In essence, they are designed to mimic electron 
motion in a 2D sample at high magnetic fields which, in the 
presence of a smoothly varying disorder potential, follows chiral
edge channels around electron puddles that are formed in local 
minima of the potential. Tunneling between these puddles occurs 
through saddle points of the disorder potential 
[Fig.~\ref{elemcell}(a)], each characterized by transmission and 
reflection amplitudes $t$ and $r$. The entire sample is then 
represented by a 2D network of four-fold vertices, connected by 
directed links. Numerical studies of a particular 
chess-board-like network\cite{chalk:jpc:88} have addressed the 
critical behavior near QH transitions. Most of the resulting 
estimates of critical properties were confirmed by real-space 
renormalization-group studies\cite{raikh:prb:97} performed 
on hierarchical networks. In the present study, we adopt a 
similar approach, but focus on the asymptotic behavior of 
transport coefficients deep in the metallic and insulating phases
rather than at the critical point. 
 
We consider a special realization of a hierarchical lattice that 
is formed by iterative replacement of a single vertex by the  
network element shown in Fig.~\ref{elemcell}(c). [The network 
resulting after the second iteration is depicted in 
Fig.~\ref{elemcell}(d).] We have chosen this particular form of 
the elementary cell in order to model a true four-terminal 
geometry that is required to measure the Hall resistance. To see 
this, let us denote its four plaquettes (corresponding to filled 
QH puddles) by symbols ${\mathcal P}_{jk}$, where $j,k$ are the 
labels of the two elementary scatterers (saddle points) on the 
respective plaquette's corners. The current is then fed into our 
elementary cell via plaquette ${\mathcal P}_{12}$ and drained 
from ${\mathcal P}_{34}$. No net current is entering or leaving 
plaquettes ${\mathcal P}_{14}$ or ${\mathcal P}_{23}$. The 
chemical potentials established in the outer links of ${\mathcal 
P}_{14}$ and ${\mathcal P}_{23}$ contain information about the 
Hall-voltage drop in the system; they could be measured by
weakly coupled voltage probes. 
 
Measurements or calculations attempting to characterize a system 
with randomness are, however, quite tricky in general. Usually, 
certain averaging procedures are involved, which may or may not 
be adequate to obtain the correct physics. In particular, large 
quantum fluctuations prevent self-averaging of physical 
quantities like, e.g., the resistance. It is then not possible to
obtain reliable results when carelessly considering averages 
only. Rather, the entire distribution function has to be taken 
into account. Alternatively, if certain self-averaging quantities
can be found, physical properties depending on these can be 
calculated reliably. A familiar example is a 1D chain of $N$ 
scatterers\cite{pwa:prb:80}. The amplitude $t_{jk}$ for 
transmission through two scatterers connected in series is given 
in terms of transmission amplitudes $t_j,t_k$ for the individual 
scatterers and a random phase $\varphi_{jk}$ \cite{fn1}: 
\begin{equation}\label{serial} 
\frac{1}{t_{jk}}=\frac{1-r_j\, r_k\, e^{i\varphi_{jk}}}{t_j\, 
t_k}\quad . 
\end{equation} 
As the (uniform) phase average of $\ln|t_{jk}|^2$ turns out to be
additive [$\langle\ln|t_{jk}|^2\rangle_{\varphi_{jk}}=\ln(t_j^2 
t_k^2)$], and higher moments of its distribution are 
bounded\cite{pwa:prb:80}, it is meaningful to define the {\em 
typical\/} transmission of a single scatterer by $T_0=\exp\{ 
\langle\ln|t_j|^2\rangle_{t_j}\}$. The typical transmission 
probability of the entire chain, given by $T_0^N$, vanishes 
exponentially with system size; i.e., the system shows {\em 
exponential localization\/}\cite{pwa:prb:80}. For a system of $N$ 
scatterers connected {\em in parallel}, the roles of transmission and 
reflection are exchanged, and the system exhibits exponential 
{\em delocalization\/}. As the connectivity of QH networks completely 
intertwines serial and parallel connection of scatterers, 
competing tendencies to localization and delocalization exist, 
leading to a localization-delocalization  transition. The 
hierarchical network considered here accounts for this aspect of 
a real QH network while, at the same time, being amenable to 
analytical study. 
 
It is straightforward to solve the transmission problem for the 
elementary cell [Fig.~\ref{elemcell}(c)]. The expression for the 
reflection probability $R^{(1)}=|o_{\text{le}}|^2/ 
|i_{\text{le}}|^2$ is 
\begin{equation} 
R^{(1)} = \frac{(1-|t_{14}|^2)(1-|t_{23}|^2)}{1 + |t_{14}|^2 
|t_{23}|^2 - 2 |t_{14}| |t_{23}| \cos(\phi - \varphi_{14} - 
\varphi_{23})}\, , 
\end{equation} 
where $t_{14}$ ($t_{23}$) is defined in terms of the transmission
amplitudes $t_j$ for scatterers labeled 1 and 4 (2 and 3), and 
the phase $\varphi_{14}$ ($\varphi_{23}$) acquired when moving 
around the plaquette ${\mathcal P}_{14}$ (${\mathcal P}_{23}$), 
according to Eq.~(\ref{serial}). Essentially, this corresponds to
a parallel connection of two scatterers with transmissions 
$|t_{14}|^2$ and $|t_{23}|^2$, each resulting from the serial 
connection of two single scatterers. The transmission probability
$T^{(1)}\equiv |o_{\text{ri}}|^2/|i_{\text{le}}|^2$ is obviously 
given by $T^{(1)}=1-R^{(1)}$. For our purposes, we also need to 
know the quantities $P^{(1)}_{14}\equiv |a_{\text{14}}|^2/ 
|i_{\text{le}}|^2$ and $P^{(1)}_{23}\equiv|a_{\text{23}}|^2 / 
|i_{\text{le}}|^2$ which represent the chemical potentials that 
would be measured by voltage probes coupled weakly to the 
outermost links of plaquettes ${\mathcal P}_{14}$ and 
${\mathcal P}_{23}$, respectively. We find 
\begin{mathletters}\label{voltages} 
\begin{eqnarray} 
P^{(1)}_{14} &=& \frac{|t_{14}|^2}{t_1^2 t_4^2}\frac{\left| t_4\,
t_{23}\, e^{i\phi} - t_1 \right|^2}{(1-|t_{14}|^2)(1-|t_{23}|^2)}
\,\, R^{(1)} \quad , \\ 
P^{(1)}_{23} &=& \frac{r_3^2}{t_3^2} \frac{|t_{23}|^2}{1 - 
|t_{23}|^2} \,\,  R^{(1)} \quad . 
\end{eqnarray} 
\end{mathletters} 
Finally, we consider phase averages for the logarithm of 
reflection and transmission probabilities for the elementary 
cell. It is straightforward to derive the exact results 
\begin{mathletters}\label{paverage} 
\begin{eqnarray}\label{tranpav} 
\langle \ln T^{(1)} \rangle_{\phi,\varphi_{14},\varphi_{23}} &=& 
\langle\ln\left[\max \left\{|t_{14}|^2,|t_{23}|^2\right\}\right] 
\rangle_{\varphi_{14},\varphi_{23}}\, , 
\\ \label{reflpav} 
\langle \ln R^{(1)} \rangle_{\phi,\varphi_{14},\varphi_{23}} &=& 
\ln R^>_{14} + \ln R^>_{23} \quad , 
\end{eqnarray} 
with $R^>_{jk}\equiv\max\{r_j^2, r_k^2\}$. In the limit of a 
broad distribution of transmissions for a single scatterer, 
Eq.~(\ref{tranpav}) yields the approximate expression 
\begin{equation}\label{tranpavapr} 
\langle \ln T^{(1)} \rangle_{\phi,\varphi_{14},\varphi_{23}} 
\approx \ln\left[\max\{t_1^2 t_4^2, t_2^2 t_3^2\}\right]\quad . 
\end{equation} 
\end{mathletters} 
We give phase averages for $\ln P^{(1)}_{jk}$ here for later use:
\begin{mathletters}\label{pavvolt} 
\begin{eqnarray} 
\langle \ln P^{(1)}_{14}\rangle_{\phi,\varphi_{14},\varphi_{23}} 
&=&\langle\ln\left[\max\{t_1^2,t_4^2\,|t_{23}|^2\}\right] 
\rangle_{\varphi_{14},\varphi_{23}}\, , \\ 
\langle \ln P^{(1)}_{23}\rangle_{\phi,\varphi_{14},\varphi_{23}} 
&=& \ln\left[t_2^2\, r_3^2\, R^>_{14}\right] \quad . 
\end{eqnarray} 
\end{mathletters} 
 
With these results, we are now in the position to derive a 
scaling theory for reflection and transmission probabilities of 
the hierarchical lattice. Successive performance of phase 
averages, starting with the 1st-level sublattices (i.e., the 
elementary cells), yields recursion relations for averages taken 
over the {\em full\/} distribution functions at the $n$th level of 
the hierarchy: 
\begin{mathletters} 
\begin{eqnarray} 
\langle \ln T^{(n)}\rangle &=& 2\langle \ln T^{(n-1)}\rangle + 
\frac{1}{2}\left\langle\left| \ln \frac{T^{(n-1)}_1 T^{(n-1)}_4} 
{T^{(n-1)}_2 T^{(n-1)}_3}\right|\right\rangle , \\ 
\langle \ln R^{(n)}\rangle &=& 2\langle \ln R^{(n-1)}\rangle + 
\left\langle\left| \ln \left[ R^{(n-1)}_1 / R^{(n-1)}_4\right] 
\right|\right\rangle . 
\end{eqnarray} 
\end{mathletters} 
Here we used Eqs.~(\ref{paverage}) and the relation $\max\{x,y\}=
(x + y + |x-y|)/2$. Assuming the logarithm of $T^{(n)}$, 
$R^{(n)}$ to be approximately normally distributed with standard 
deviations $\sigma_T^{(n)}$, $\sigma_R^{(n)}$, we obtain for 
$n\gg 1$ 
\begin{mathletters} 
\begin{eqnarray} 
\langle \ln T^{(n)}\rangle &=& 2\langle \ln T^{(n-1)}\rangle + 
\sqrt{\frac{2}{\pi}} \,\sigma^{(n-1)} \quad , \\ 
\langle \ln R^{(n)}\rangle &=& 2\langle \ln R^{(n-1)}\rangle + 
\frac{2}{\sqrt{\pi}}\,\sigma^{(n-1)} \quad , \\ 
\sigma^{(n)}_{T(R)} &=& \alpha^n\,\sigma_{T(R)} \quad .  
\end{eqnarray} 
\end{mathletters} 
Here, $\alpha=\sqrt{2(\pi-1)/\pi}=1.1676\dots$; $\sigma_{T(R)}^2$
is {\em weakly\/} dependent on $n$ via $\sigma_{T(R)}^2\equiv  
[\sigma_{T(R)}^{(0)}]^2 + \Delta_{T(R)}^2/(\alpha^2-1)$, where 
$\sigma_{T(R)}^{(0)}$ denotes the standard deviation for the 
distribution of $\ln T^{(0)}$ ($\ln R^{(0)}$) associated with the
elementary scatterers in our lattice, and $\Delta_{T}\leq \pi$
($\Delta_{R}\leq\sqrt {5/3}\,\pi$) 
results from phase averaging of $\ln [T^{(n)}]^2$ 
($\ln [R^{(n)}]^2$). The recursion relations are easily converted
into explicit scaling expressions for the {\em typical\/} $T^{(n)}$,
$R^{(n)}$: 
\begin{mathletters}\label{scaling} 
\begin{eqnarray} 
T^{(n)}_{\text{typ}} &=& \exp\{\langle\ln T^{(n)}\rangle\} = 
\left({\tilde T}_0\right)^{2^n} \quad , \\ 
R^{(n)}_{\text{typ}} &=& \exp\{\langle\ln R^{(n)}\rangle\} = 
\left({\tilde R}_0\right)^{2^n} \quad , 
\end{eqnarray} 
\end{mathletters} 
with the abbreviations 
\begin{mathletters} 
\begin{eqnarray} 
{\tilde T}_0 &=& T_0 \,\exp\left\{\frac{\sqrt{2}}{\sqrt{\pi}(2- 
\alpha)}\,\sigma_T\right\} \quad , \\ 
{\tilde R}_0 &=& R_0\, \exp\left\{\frac{2}{\sqrt{\pi}(2-\alpha)} 
\,\sigma_R\right\} \quad . 
\end{eqnarray} 
\end{mathletters} 
The quantities $T_0,R_0$ are the usual typical transmission and 
reflection probabilities of elementary scatterers. Considering 
the ratio of average and standard deviation for $\ln R^{(n)}$, 
we find 
\begin{equation} 
\frac{\sigma_R^{(n)}}{|\langle\ln R^{(n)}\rangle|}=\left( 
\frac{\alpha}{2}\right)^n \, \frac{\sigma_R}{|\ln{\tilde R}_0|} 
\quad , 
\end{equation} 
which scales to zero in the limit of large $n$. The same relation
holds for the transmission. Hence, the typical scaling behavior 
obtained in Eqs.~(\ref{scaling}) is indeed representative of the 
behavior of the full distributions of reflection and transmission
probabilities of a large hierarchical lattice. Note that, in 
contrast to any 1D situation, the typical values of {\em both\/} 
reflection and transmission probabilities for our network scale 
to zero exponentially with system size. This is a signature of 
the coexisting trends toward localization and delocalization in 
our network. 
 
Physical quantities whose logarithmic average reduces to that of 
reflection and transmission probabilities will exhibit meaningful
typical averages as well. This is obviously the case for the 
dimensionless longitudinal resistance of the $n$th-level 
hierarchical lattice, which is given by $\rho_{\text{L}}^{(n)} = 
R^{(n)}/T^{(n)}$. Using Eqs.~(\ref{scaling}), we find 
$\rho_{\text{L,typ}}^{(n)} = \left({\tilde R}_0/{\tilde T}_0 
\right)^{2^n}$. This result implies that there exist two phases, 
a metallic one for ${\tilde R}_0<{\tilde T}_0$, and an insulating
one for ${\tilde R}_0>{\tilde T}_0$. The unstable critical point 
separating the two is defined by ${\tilde R}_0={\tilde T}_0$. 
Hence, the phenomenology of the localization-delocalization 
transition is reproduced, similarly to other QH network 
models\cite{chalk:jpc:88,raikh:prb:97}. 
 
We now turn to the discussion of the Hall resistance. Deep in the
QH phases where $\rho_{\text{L}}\approx 0$, it can be measured
directly in a four-terminal geometry \cite{butt:prb:88}. In
general, however, this four-terminal voltage $V_{\text{4t}}$
measures a combination of both longitudinal and Hall-voltage
drop. To disentangle the two contributions, one uses the fact
that the Hall voltage is antisymmetric under reversal of the
direction of the magnetic field $B$, and defines it via 
$V_{\text{H}} =[V_{\text{4t}}(B)-V_{\text{4t}}(-B)]/2$. Applying 
this concept to our situation, we define the (normalized) Hall 
voltage of the elementary cell by 
\begin{equation} 
V_{\text{H}}^{(1)}=\frac{1}{2}\left[\left(P^{(1)}_{14}- 
P^{(1)}_{23}\right)_B - \left(P^{(1)}_{14}-P^{(1)}_{23} 
\right)_{-B} \right] \quad . 
\end{equation} 
In a QH network, field reversal corresponds to reversal of the
propagation direction along the chiral edge channels (represented
by arrows in Fig.~\ref{elemcell}). Transmission and reflection
probabilities are, of course, unaffected by this operation. The
explicit expression for the Hall resistance
$\rho_{\text{H}}^{(1)}=V_{\text{H}}^{(1)}/T^{(1)}$ of the 
elementary cell can be written as 
\widetext 
\top{-2.8cm} 
\begin{equation}\label{flucthall} 
\rho_{\text{H}}^{(1)}-1 = \frac{1}{2}\frac{\sum_{j=1}^4 t_j^{-2} 
-2-|t_{14}|^{-2}-|t_{23}|^{-2}-\left[e^{i\phi}\left(\frac{r_1 
r_4}{t_1 t_4}\frac{e^{-i\varphi_{14}}}{t_{23}^*} + \frac{r_2 r_3}
{t_2 t_3} \frac{e^{-i\varphi_{23}}}{t_{14}^*}\right)+\mbox{c.c.} 
\right]}{|t_{14}|^{-2}+|t_{23}|^{-2} - 2 \cos(\phi - \varphi_{14}
-\varphi_{23})/|t_{14}||t_{23}|}\; . 
\end{equation} 
Generalization to the hierarchical network is straightforward.  
Inspection shows that $\rho_{\text{H}}^{(n)}-1={\mathcal O}\left(
R^{(n-1)}\right)\ll 1$ in the metallic regime. In the insulator, 
however, $\rho_{\text{H}}^{(n)}$ fluctuates strongly, and the 
expression given in Eq.~(\ref{flucthall}) does not yield a 
physically meaningful value. Instead, a typical Hall resistance 
has to be considered which we define by 
\begin{equation} 
\rho_{\text{H,typ}}^{(n)}=\frac{1}{2}\left[\left.\frac{\exp\{ 
\langle\ln P_{14}^{(n)}\rangle\}-\exp\{\langle\ln P_{23}^{(n)} 
\rangle\}}{\exp\{\langle\ln T^{(n)}\rangle\}}\right|_B - \left. 
\frac{\exp\{\langle\ln P_{14}^{(n)}\rangle\}-\exp\{\langle \ln 
P_{23}^{(n)}\rangle\}}{\exp\{\langle\ln T^{(n)}\rangle\}} 
\right|_{-B} \right] \quad . 
\end{equation} 
\narrowtext 
\noindent 
The quantities $P_{14}^{(n)}$ and $P_{23}^{(n)}$ are given by 
generalization of Eqs.~(\ref{voltages}) to the case of the 
hierarchical network after $n$ iterations. Their values for the 
field-reversed situation can be obtained by making the change 
$1\leftrightarrow 2$ and $3\leftrightarrow 4$ in all labels. 
Using phase averages derived earlier [Eqs.~(\ref{pavvolt})], we 
relate the fluctuating chemical potentials to the 
transmission/reflection probabilities via 
\begin{mathletters} 
\begin{eqnarray} 
\langle\ln P_{14}^{(n)}\rangle &=& \langle\ln T^{(n-1)}\rangle 
\quad ,\\ 
\langle\ln P_{23}^{(n)}\rangle &=& \langle\ln T^{(n-1)}\rangle + 
2 \langle\ln R^{(n-1)}\rangle+\frac{\sigma^{(n-1)}}{\sqrt{\pi}} 
\, . 
\end{eqnarray} 
\end{mathletters} 
These translate into an explicit scaling expression for the 
typical Hall resistance, 
\begin{equation}\label{hallres} 
\rho_{\text{H,typ}}^{(n)}=\left[1-\left({\tilde R}_0\right)^{2^n}
\right]/\left({\tilde T}_0\right)^{2^{n-1}}\approx \left({\tilde 
T}_0\right)^{-2^{n-1}} , 
\end{equation} 
which scales to infinity {\em exponentially\/} with system size 
$L=2^n$. Equation~(\ref{hallres}) supports numerical  
evidence\cite{assa:prl:99} that the typical Hall resistance of a 
phase-coherent QH system {\em diverges\/} in the insulating regime, 
though more moderately than $\rho_{\text{L,typ}}$. Note that in 
the strongly insulating limit (${\tilde T}_0\ll {\tilde R}_0$), 
our hierarchical model yields a scaling relation 
$\rho_{\text{H,typ}}\propto\left(\rho_{\text{L,typ}} 
\right)^\gamma$ with $\gamma\approx 1/2$. The corresponding 
localization lengths $\xi_L$, $\xi_H$ (defined via 
$\rho_{\text{L(H),typ}}\sim \exp\{L/\xi_{L(H)}\}$) are related 
by $\xi_H\approx 2\xi_L$. 
 
In conclusion, we have derived a scaling theory for both 
longitudinal and Hall resistivities in a hierarchical network 
model of a QH system. A transition from a localized (insulating) 
phase to a delocalized (metallic) phase is found. In agreement 
with previous numerical results, we find that the Hall 
resistivity diverges in the insulating regime as a consequence of
phase coherence of electron transport in the network. Hence, we 
expect the characteristics of a QHI to occur only in a regime
where electrons dephase between elastic scattering events.
 
We thank A. Auerbach, O. Entin-Wohlmann, F. Evers, Y. Gefen, 
B. Huckestein, M. Janssen, A. MacKinnon, A. Mirlin, D. Polyakov,
L. Pryadko, B. Shapiro, D. Sheng, S. Sondhi, and I. Zharekeshev
for useful discussions. This work was supported by the DIP
project of the German Ministry for Education and Research (BMBF),
Grant No.\ 96--00294 from the United States--Israel Binational
Science Foundation (BSF), and the Aspen Center for Physics. U.Z.\
thanks the Institute of Theoretical Physics at the Technion ---
Israel Institute of Technology for hospitality during two visits
when part of this research was performed.  
 

\begin{thebibliography}{10}
\bibitem{fn1} 
Note that $t_j$, $r_j$ are real. Phase shifts associated with 
single scatterers are absorbed in the definition of 
$\varphi_{jk}$.
\end{thebibliography}

\widetext 

\end{document}